\documentclass[aps,pre,superscriptaddress,twocolumn,amsmath,amssymb,showpacs]{revtex4}

\usepackage{graphicx}
\usepackage{bm}
\newcommand{\ignore}[1]{} \usepackage{epsfig} \usepackage{color}
\usepackage{amsmath,amssymb}

\begin{document}

\title{Rescue of endemic states in interconnected networks with
  adaptive coupling}

\author{F. Vazquez} \affiliation{IFLYSIB, Instituto de F\'isica de
  L\'iquidos y Sistemas Biol\'ogicos (UNLP-CONICET), 1900 La Plata,
  Argentina}
\email[E-mail: ]{fede.vazmin@gmail.com}

\author{M. \'Angeles Serrano} \affiliation{Departament de F\'isica
  Fonamental, Universitat de Barcelona, Mart\'i i Franqu\`es 1, 08028,
  Barcelona, Spain} \affiliation{Instituci\'o Catalana de Recerca i
  Estudis Avan\c{c}ats (ICREA), Barcelona 08010, Spain}

\author{M. San Miguel} \affiliation{IFISC, Instituto de F\'isica
  Interdisciplinar y Sistemas Complejos (CSIC-UIB), E-07122 Palma de
  Mallorca, Spain}

\date{\today}

\begin{abstract}
We study the Susceptible-Infected-Susceptible model of epidemic
spreading on two layers of networks interconnected by adaptive links, which are
rewired at random to avoid contacts between infected and susceptible
nodes at the interlayer.  We find that the rewiring reduces the effective
connectivity for the transmission of the disease between layers, and
may even totally decouple the networks.  Weak endemic states, in which
the epidemics spreads only if the two layers are interconnected, show a
transition from the endemic to the healthy phase when the rewiring
overcomes a threshold value that depends on the infection rate, the
strength of the coupling and the mean connectivity of the networks. In
the strong endemic scenario, in which the epidemics is able to spread
on each separate network, the prevalence in each layer decreases when
increasing the rewiring, arriving to single network values only in the
limit of infinitely fast rewiring.  We also find that finite-size
effects are amplified by the rewiring, as there is a finite
probability that the epidemics stays confined in only one network
during its lifetime.  
\end{abstract}

\pacs{89.75.Hc, 05.45.Df, 64.60.Ak}

\maketitle

\section{Introduction}

Most biological and sociotecnological systems in nature are not isolated, but
they are formed by a set of complex
networks which interact following intricate patterns. These systems
can be represented in terms of a multilayer
network~\cite{Domenico:2013,Kivela:2014} with interconnected layers,
in which each network layer describes a different type of
connectivity, process, or population. The function of these
structures, like sustaining diffusion processes, is characterized by
emerging features which are not observed in single-layered
systems~\cite{Domenico:2015,Diakonova:2015}. In particular, epidemics
propagating on two interconnected networks show a direct dependence on
the patterns of interconnection between the layers. Previous work on
the Susceptible-Infected-Susceptible (SIS) model~\cite{Saumell:2012}
showed that interconnected systems can support an endemic state even
if each isolated network is not able to do so, and that internetwork
correlations boost the propagation on the two-network system.  In a
related work \cite{Dickison:2012} the authors found
a mixed phase in the Susceptible-Infected-Recovered (SIR) dynamics
--besides the endemic and the healthy phase observed in isolated
networks--, in which the epidemics spreads in one network only.  The SIR and SIS
dynamics on multiplexes was also explored in some recent papers
\cite{Buono:2014,Granell:2014,Sanz:2014}.

These and many other works on multilayered systems considered static
links inside and between layers, since contacts between individuals
are assumed to be constant over time.  However, it is known that the
frequency and duration of face-to-face contacts are quite
heterogeneous, and follow non-trivial patterns \cite{Barrat}.  Past
studies in single networks incorporated different mechanisms of time
varying topology and network adaptation \cite{Holme:2013}.  A prominent
example of time dependent networks are coevolving networks, in which
the state of each node is influenced by, and simultaneously shapes, the
local network structure
\cite{Zimmermann:2001,Zimmermann:2004,Gross:2008,Gross:2009,Vazquez:2013}.  This
coupling of the dynamics \emph{on} the network with the dynamics
\emph{of} the network  leads to a rich variety of different long term
behavior of the system. Simple contact processes like the SIS model
\cite{Gross:2006}, the voter model
\cite{Vazquez:2008sd,Diakonova:2015b} or the Axelrod model
\cite{VazquezAvella:2007} on coevolving networks  give rise to new
phenomena, not seen in static networks, such as self-organization
towards critical behavior, the formation of complex topologies and
network fragmentation transitions.  Recently, coevolution in mutilayared
networks has considered the dynamics of links inside each layer
or intralayer links \cite{Shai:2013,Diakonova:2014}, but little
attention has been paid to the coevolution of links between layers
(interlayer links).

In this paper, we study the SIS model of epidemic spreading on
interconnected networks with dynamic connections between layers.
Links connecting susceptible and infected individuals at the
interlayer are rewired and reconnected to a random pair of
susceptible individuals.  In this way the dynamics of interlayer links is
coupled to the dynamical evolution of the states of the nodes.  A
related work \cite{Shai:2013}
considered this adaptive mechanism, but only at the level of
intra-network connections, keeping interconnections static.  We find
that the network adaptation process associated with the
rewiring of interlayer links can dramatically reduce the spreading of
the disease within and between layers.  We consider two different scenarios. In
weak endemic
states, when the epidemics spreads only if the two layers are
interconnected, we find a critical rewiring rate beyond which the
system remains in the healthy phase.  For strong endemic states, in
which the epidemics is able to spread on each separate network, we
find that rewiring is specially effective in preventing spreading on
finite systems.  This is so because the link adaptation amplifies
finite-size effects, leading to a finite probability that the
epidemics stays confined in only one layer during its lifetime.

The paper is organized as follows.  In section \ref{model}, we
introduce the model and describe the multilayer topology and the
dynamics.  In section \ref{math}, we develop the mathematical
framework to study the dynamical properties of the system under two
different scenarios, weak endemic states in section \ref{weak} and
strong endemic states in section \ref{strong}.  Also, in section
\ref{strong} we study the likelihood of transmission between layers in
finite systems.  We summarize our results and give conclusions in
section \ref{conclusions}.

\section{Model description}
\label{model}

We study a system of two coupled networks or layers, connected by
adaptive interconnections.  To build the topology of the system we
start from two Erd\"os-Renyi networks A and B which, for the sake of
simplicity, have the same
number of nodes $N_A=N_B=N$ and average degrees $\langle k_A \rangle =
\langle k_B \rangle = \langle k \rangle$.  Then, these networks are
interconnected with links placed at random between nodes in A and B,
whose number remains constant over time, so that the total number of
links in the system is conserved.  The number of
inter-links $q \langle k \rangle N/2$ is proportional to the number of
intra-links $\langle k \rangle N/2$ inside each network, where the
factor $q \ge 0$ is a measure of the intensity of the coupling, so that
for $q=0$ the networks are totally decoupled.  As a result, nodes in
each network have an average number $\langle k \rangle$ of
intranetworks connections and an average number $\langle k_{AB}
\rangle=q \langle k \rangle/2$ of internetworks connections.

An SIS epidemics spreads on top of the coupled system, thus each node
can be in one of two possible states, either susceptible (state $S$) or
infected (state $I$).  Each infected node transmits the disease to its
susceptible neighbors at some infection rate $\lambda$ ($\lambda_A$
and $\lambda_B$ inside each layer, and $\lambda_{AB}$ ($\lambda_{BA}$)
from nodes in layer A (B) to nodes in layer B (A)), and recovers
becoming susceptible again at rate $\mu$, that without loss of
generality we take as $\mu=1$.  Intralayer links in A and B are fixed
in time, but interlayer
connections between A and B are allowed to adapt over time. That is,
every interconnection between an infected node in network A (B) and a
susceptible node in B (A) is randomly rewired at rate a $\omega$. To
implement the rewiring, the selected Infected-Susceptible ($IS$)
interlink is removed and a new Susceptible-Susceptible ($SS$)
interlink is created in its place, in which the susceptible nodes are
chosen at random in different layers. In the continuous time version
of the model, at
every infinitesimal time step $dt$ the processes of infection,
recovery and rewiring happen with probabilities $\lambda dt$, $dt$ and
$\omega dt$, respectively.  We use the Gillespie
method~\cite{Gillespie:2001kl} to simulate the dynamics. At every
iteration step of discrete time length $\Delta t=1/R$, one of the
three processes is performed with respective probabilities
$\lambda/R$, $1/R$ and $\omega/R$, where $R=\lambda+1+w$ is the total
rate.

\section{Mathematical framework}
\label{math}

We develop a mathematical approach in the thermodynamic limit of
infinitely large systems, based on a mean-field description at the
level of pairs of nodes.  This approach allows to study the dynamics of the
system in terms of the
time evolution of the global densities of nodes and links in the
different states. We denote by $I_A$ and $S_A$ the densities of
infected and susceptible nodes in network A, respectively, and by
$I_AS_A$, $I_AI_A$
and $S_AS_A$ the densities per node of intranetwork links in A
connecting an infected and a susceptible node, two infected nodes and
two susceptible nodes, respectively.  An analogous notation is used
for network B.  For interconnections, we use the notation $I_AS_B$
($S_AI_B$) to represent the density per node of internetwork links
between an infected node in A (B) and susceptible node in B (A), while
$I_AI_B$ and $S_AS_B$ stand for the densities of interlinks between
two infected nodes and two susceptible nodes, respectively.  We note
that these quantities are symmetric with respect to the subindices A
and B.

Given that the number of nodes in each layer is conserved, as well as
the number of intra- and interlinks, the following conservation
relations hold at any time:
\begin{subequations}
\label{conserv}
\begin{eqnarray}
1 &=& I_A+S_A, \\ 1 &=& I_B+S_B, \\ \frac{\left<k_A\right>}{2} &=&
S_AS_A+I_AI_A+S_AI_A, \\ \frac{\left<k_B\right>}{2} &=&
S_BS_B+I_BI_B+S_BI_B, \\
\label{conservAB}
\left<k_{AB}\right> &=& S_AS_B+I_AS_B+S_AI_B+I_AI_B.
\end{eqnarray}
\end{subequations}

In order to obtain a closed system of equations for the evolution of
the densities we make use of the
pair approximation (see Appendix \ref{pair}), which assumes that the
links of different types are homogeneously distributed over the
networks, and thus the densities of triplets can be written in terms
of the densities of pairs.  We obtain
\begin{subequations}
\label{dIdtpa}
\begin{eqnarray}
\label{dIAdt}
\frac{dI_A}{dt}&=&-  I_A + \lambda_A I_AS_A + \lambda_{BA} S_AI_B, \\
\label{dIASAdt}
\frac{dI_AS_A}{dt} &=& 2I_AI_A - (1+\lambda_A) I_AS_A - \lambda_A
\frac{I_AS_A \cdot I_AS_A}{S_A} \nonumber \\ &-& \lambda_{BA}
\frac{I_AS_A \cdot S_AI_B}{S_A} + 2 \lambda_A \frac{S_AS_A \cdot
  I_AS_A}{S_A} \nonumber \\ &+& 2 \lambda_{BA} \frac{S_AS_A \cdot
  S_AI_B}{S_A}, \\
\label{dIAIAdt}
\frac{dI_AI_A}{dt}&=&- 2 I_AI_A + \lambda_A I_AS_A  +  \lambda_A
\frac{I_AS_A \cdot I_AS_A}{S_A}  \nonumber \\ &+& \lambda_{BA}
\frac{I_AS_A\cdot S_AI_B}{S_A}, \\
\label{dIASBdt}
\frac{dI_AS_B}{dt} &=& I_AI_B - (1 + \omega) I_AS_B -  \lambda_{AB}
I_AS_B \nonumber \\ &+& \lambda_A \frac{I_AS_A \cdot S_AS_B}{S_A}  -
\lambda_B \frac{I_BS_B \cdot I_AS_B}{S_B} \nonumber \\ &-&
\lambda_{AB} \frac{I_AS_B \cdot I_AS_B}{S_B} +
\lambda_{BA}\frac{S_AI_B \cdot S_AS_B}{S_A}, \\
\label{dSAIBdt}
\frac{dS_AI_B}{dt} &=& I_AI_B - (1 + \omega) S_AI_B -  \lambda_{BA}
S_AI_B \nonumber \\ &+& \lambda_B \frac{S_AS_B \cdot I_BS_B}{S_B}  -
\lambda_A \frac{I_AS_A \cdot S_AI_B}{S_A} \nonumber \\ &-&
\lambda_{BA} \frac{S_AI_B \cdot S_AI_B}{S_A} +
\lambda_{AB}\frac{I_AS_B \cdot S_AS_B}{S_B}, \\
\label{dIAIBdt}
\frac{dI_AI_B}{dt}&=&- 2  I_AI_B + \lambda_{AB} I_AS_B + \lambda_{BA}
S_AI_B \nonumber \\ &+& \lambda_A \frac{I_AS_A \cdot S_AI_B}{S_A}  +
\lambda_B \frac{I_BS_B \cdot I_AS_B}{S_B} \nonumber \\ &+&
\lambda_{AB} \frac{I_AS_B \cdot I_AS_B}{S_B} +  \lambda_{BA}
\frac{S_AI_B \cdot S_AI_B}{S_A}.
\end{eqnarray}
\end{subequations}
The pair approximation works reasonably well for networks with
homogeneous degree distributions and low degree correlations, such as
Erd\"os-R\'enyi (ER) or degree-regular random graphs, but may not be
accurate enough for non-homogeneous or correlated toplogies, like
scale-free networks.  Even for ER networks, quantitative disagreements
between the simulations and the analytical solution may arise in some
regimes due to dynamic correlations not captured by the approximation,
as it happens when the system is close to the healthy absorbing state
\cite{Demirel:2014}.

The set of non-linear coupled ordinary differential
Eqs.~(\ref{dIdtpa}) together with the conservation laws
Eqs.~(\ref{conserv}) form a closed set of equations for the system's
dynamics.  This mathematical framework is suitable to study
macroscopic quantities, such as the prevalence of the epidemics
(stationary density of infected nodes $I_A$ and $I_B$) in each layer.  We are
particularly interested in the effects of the rewiring on the
prevalence, thus we focus on the dependence of $I_A$ and $I_B$ on the
rewiring rate $\omega$.  We shall see that the behavior of the coupled
system at the stationary state differs qualitatively depending on
whether the global endemic state is \emph{weak} or \emph{strong}, with
strong endemic states having the capacity to propagate on each network
separately, while weak endemic states are unable to do
so~\cite{Saumell:2012}. We recall that strong endemic states are
supported by infectivity levels above the critical value $\lambda_c$
in each single layer, while weak endemic states appear whenever the
infectivity is below the single epidemic threshold $\lambda_c$, but
above the epidemic threshold $\lambda_{c,q}$ of the interconnected
system, which depends on the coupling $q$.  In single random
uncorrelated networks the critical threshold is typically given by the
expression $\lambda_c = \left< k \right>/\left< k^2 \right>$
\cite{Pastor-Satorras:2001fl}, and it can be better approximated by
$\lambda_c = \left< k \right> / \left< k(k-1) \right>$
\cite{Saumell:2012}, which corrects in part the effect of dynamical
correlations.  In ER networks $\left< k^2 \right> = \left< k \right>^2
+ \left< k \right>$, and thus the single-layer threshold is reduced to
the simple expression \cite{Luo:2014}
\begin{equation}
\lambda_c= \frac{1}{\left<k\right>}.
\label{lambdac}
\end{equation}

\section{Weak endemic states}
\label{weak}

As we mentioned above, the existence of endemic states in the coupled
system of two layers is possible even when infection rates are below
the critical rate $\lambda_c$ of a single layer, as was shown in
\cite{Saumell:2012}.  It turns out that the infection in each layer is
reinforced by its partner layer through the interconnections, and thus
the entire system lowers its epidemic threshold to a new value
$\lambda_{c,q} < \lambda_c$, associated to the interconnectivity $q$.
Therefore, endemic sates are also observed for infection rates in the
region $\lambda_{c,q} \le \lambda \le \lambda_c$.  In this section we
focus on the $\lambda < \lambda_c$ case, and analyze separately the
case scenarios of static and dynamic interconnections.

\subsection{Static interlayer links}

\begin{figure}[t]
\includegraphics[width=7.5cm]{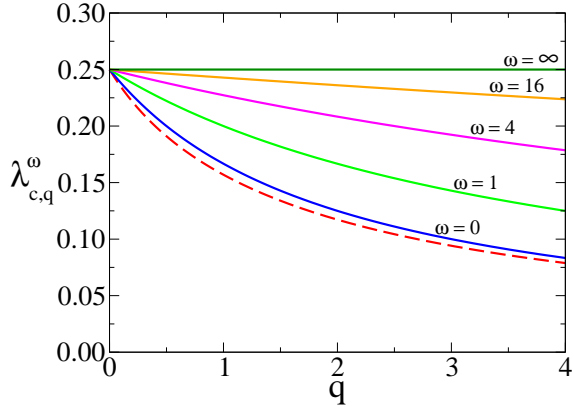}
\caption{Critical infection rate $\lambda_{c,q}^\omega$ of a system of
  two interconnected networks of average degree $\left< k \right>=4$
  vs the coupling factor $q$.  Solid lines represent the expression
  Eq.~(\ref{lambda-cqw}) for different values of the rewiring rate
  $\omega=\infty, 16, 4, 1$ and $0$ (from top to bottom), while the
  dashed line is the approximation from Eq.~(\ref{lambdacq+-}) for
  $\omega=0$}.
\label{lambdacq}
\end{figure}

For the sake of simplicity, we consider the symmetric case $\lambda_A
= \lambda_B = \lambda_{AB} = \lambda_{BA} = \lambda < \lambda_c$.
According to the general heterogeneous mean-field approach
in~\cite{Saumell:2012}, the epidemics can propagate on a system of two
identical networks whenever the dynamical and topological
characteristic parameters fulfill the condition
\begin{equation}
\alpha > (1-\Lambda)(1-\Omega).
\label{alpha}
\end{equation}
Here $\Lambda=\lambda \left<k^2\right>/\left<k\right>$ is the maximum
eigenvalue of the characteristic matrix of each single network, which
determines the condition for the existence of an endemic state
whenever $\Lambda>1$. We next particularize for the case of ER
networks. Taking into account the partial correction for the effect of
dynamical correlations~\cite{Saumell:2012}, the maximum eigenvalue for
an isolated ER network is
$\Lambda=\lambda\left<k(k-1)\right>/\left<k\right> = \lambda \left< k
\right>$.  The factor $\Omega=\lambda\left<k^2_{AB}\right>/ \left<
k_{AB} \right>$ is the counterpart for the bipartite network of
interconnections, whose first and second degree moments are $\left<
k_{AB} \right>$ and $\left< k_{AB}^2 \right>=\left< k_{AB} \right>
\left(\left< k_{AB} \right>+1 \right)$ and hence $\Omega = \lambda
\left( \left< k_{AB} \right> +1 \right)$.  Finally, the factor
$\alpha$ given by $\lambda^2\left<k \cdot
k_{AB}\right>^2/\left<k\right> \left< k_{AB} \right>$ can be
approximated as $\alpha \simeq \lambda^2 \left<k \right> \left< k_{AB}
\right>$.  Plugging the values for $\Lambda$, $\Omega$ and $\alpha$
into Eq.~(\ref{alpha}) we arrive to the condition
\begin{eqnarray}
(\lambda-\lambda_{c,q,+})(\lambda-\lambda_{c,q,-})  < 0,
\label{condition}
\end{eqnarray}
with
\begin{eqnarray}
\lambda_{c,q,\pm}=\frac{1+(1+\frac{q}{2}) \left<k\right> \pm
  \sqrt{\left[ 1+ (1+\frac{q}{2})\left<k\right>
      \right]^2-4\left<k\right>}}{2\left<k\right>}. \nonumber \\
\label{lambdacq+-}
\end{eqnarray}
Infection rates in the range $\lambda_{c,q,-}< \lambda <
\min(\lambda_{c,q,+},\lambda_c)$ fulfill the condition
Eq.~(\ref{condition}), and thus they are able to spread the epidemics
on the coupled system, even if $\lambda$ is below the single network
threshold $\lambda_c=1/\left< k \right>$.  Even though
Eq.~(\ref{lambdacq+-}) gives a good estimation of the infection
threshold, a more precise expression for $\lambda_{c,q}$ can be
obtained using the homogeneous pair approximation of section
\ref{math} [from Eqs.~(\ref{conserv}) and (\ref{dIdtpa})]. The
limit for $\omega=0$ of the general result reported in section
\ref{dynamic} is
\begin{equation}
\lambda_{c,q} = \frac{1}{(1+q/2)\left< k \right>},
\label{lambda_cq}
\end{equation}
which reduces to Eq.~(\ref{lambdac}) in the absence of coupling
$q=0$. This result correspond to the value of the critical infection
rate for a single layer ER network with average degree $\left< k
\right> + \left< k_{AB} \right> = (1+q/2)\left< k \right>$. Critical
infection rates from Eq.~(\ref{lambdacq+-}) and Eq.~(\ref{lambda_cq})
turn out to be numerically very similar [see Fig.~(\ref{lambdacq})].

\subsection{Adaptive interlayer links}
\label{dynamic}

Next, we study the behavior of the system when interconnections are
rewired at a given rate $\omega$ to avoid contacts between infected
and susceptible nodes at the interlayer.

\begin{figure}[t]
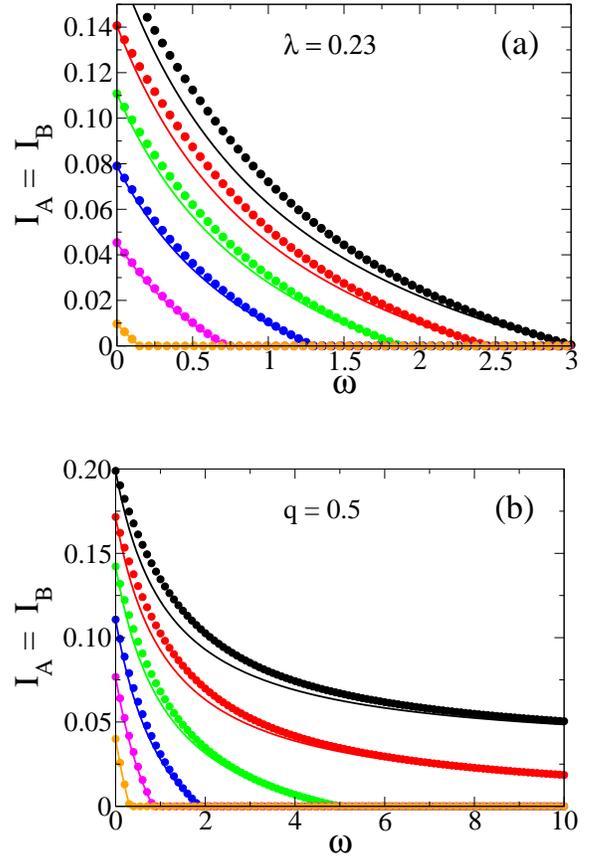

\begin{tabular}{l}
\includegraphics[width=7.5cm]{Fig2a.eps} \\\\\\
\includegraphics[width=7.5cm]{Fig2b.eps}
\end{tabular}
\caption{Disease prevalence $I_A=I_B$ in a symmetric two-network
  system as a function of the rewiring rate $\omega$.  Each network
  has average degree  $\left<k\right>=4$.  Circles are the results
  from the numerical integration of Eqs.~(\ref{dIdtpa}), while lines
  are the analytical approximation Eq.~(\ref{Isw}).  (a) For infection
  rate $\lambda = 0.23$ and values of the coupling $q=0.7, 0.6, 0.5,
  0.4, 0.3$ and $0.2$ (from top to bottom).  (b) For $q=0.5$ and
  values of the infection rate $\lambda=0.26, 0.25, 0.24, 0.23, 0.22$
  and $0.21$ (top to bottom).}
\label{fig:I-w}
\end{figure}

\begin{figure}[t]
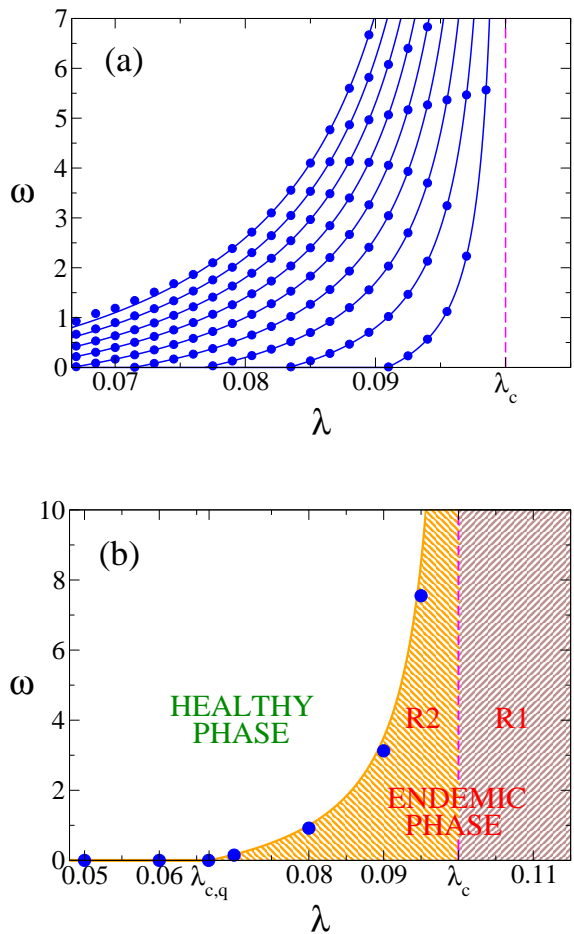

\begin{tabular}{l}
\includegraphics[width=7.5cm]{Fig3a.eps} \\\\\\
\includegraphics[width=7.5cm]{Fig3b.eps}
\end{tabular}
\caption{(a) Critical rewiring rate $\omega_{c,q}^{\lambda}$ vs
  infectivity $\lambda$ for a symmetric two-network system with
  average degree $\left< k \right>=10$, obtained from
  Eqs.~(\ref{dIdtpa}) (dots) and from Eq.~(\ref{wcqlambda}) (solid
  lines), for values of the coupling $q=0.2, 0.4, 0.6, 0.8, 1.0, 1.2,
  1.4, 1.6$ and $1.8$ (from left to right). (b) Healthy-endemic phase
  diagram for two coupled ER networks of $N=10^4$ nodes each, average
  degree $\left< k \right>=10$, and coupling $q=1.0$.  One-layer and
  two-layer critical infection rates are $\lambda_c=0.1$ and
  $\lambda_{c,q}=0.067$, respectively.  Filled circles are results
  from simulations, while the solid line is the analytical solution
  Eq.~(\ref{wcqlambda}) of the system of Eqs.~(\ref{dIdtpa}).  }
\label{fig:wc-lambda}
\end{figure}

In Fig.~\ref{fig:I-w} we show the prevalence $I_A=I_B$ in networks A
and B for a symmetric system with $\left< k \right>=4$, as predicted
by the analytic Eqs.~(\ref{dIdtpa}) (solid circles).  We show as well
the approximate analytic solution of Eqs.~(\ref{dIdtpa}) (solid lines)
that we derive in Appendix \ref{prev}.  We observe that the prevalence
decreases as the rewiring increases, and vanishes at a finite value
$\omega_{c,q}^{\lambda}$ when $\lambda < \lambda_c=1/\left< k
\right>=0.25$.  That is, there is a continuous transition from an
endemic to a healthy phase when the rewiring overcomes a critical
value $\omega_{c,q}^{\lambda}$, which depends on the coupling $q$ and
the infection rate $\lambda < 0.25$. In Fig.~\ref{fig:wc-lambda}(a) we
plot the critical rewiring rate vs the infectivity $\lambda$ for
various values of $q$ and $\left< k \right>=10$, obtained from the
numerical integration of Eqs.~(\ref{dIdtpa}) (dots).  As we can see,
$\omega_{c,q}^{\lambda}$ is larger for larger values of $q$ and
$\lambda$, and diverges as $\lambda$ approaches $\lambda_c=1/\left< k
\right>=0.1$.  This divergence means that for strong endemic states
($\lambda > 0.1$) there is no finite rewiring able to stop the
spreading.  In the next section, we explore this phenomenon in more
detail.

We now derive an analytical expression for the dependence of the
critical rewiring with the infectivity $\lambda$ and the average
connectivities $\left<k\right>$ and $q$.  To that end, we use the
system of Eqs.~(\ref{dIdtpa}) to study the stability of the healthy
phase under a small perturbation that consists on infecting a small
fraction of nodes in both layers.  For simplicity, we consider the
symmetric case scenario in which both perturbations are the same, so
that initially $I_A = I_B \ll 1$.  Because of the symmetries of the
system and initial conditions, the prevalence in both networks must be
the same, thus $I_BS_B=I_AS_A$, $I_BI_B=I_AI_A$ and $S_AI_B=I_AS_B$.
This reduces Eqs.~(\ref{dIdtpa}) to a system of five equations with
five unknown densities: $I_A, I_AS_A, I_AI_A, I_AS_B$ and $I_AI_B$.
Given that these densities depend on the number of infected nodes and
$I_A \ll 1$, they are also very small initially.  Then, we linearize
the equations around the fixed point $S_A=S_B=1, S_AS_A=S_BS_B=\langle
k \rangle/2, S_AS_B=\langle k_{AB} \rangle=q \langle k \rangle/2$, by
neglecting terms of order $2$ or higher in the five densities.  We
obtain a reduced system which can be written in matrix representation
as
\begin{equation}
\frac{d{\bf I}}{dt} = {\bf A I},
\end{equation}
where
\begin{eqnarray*}
{\bf A} \equiv \left( \begin{array}{ccccc} -1 & \lambda & 0 & \lambda
  & 0 \\ 0 & \lambda \langle k \rangle - \lambda -1 & 2 & \lambda
  \langle k \rangle & 0 \\ 0 & \lambda & -2 & 0 & 0 \\ 0 & \lambda
  \langle k \rangle & 0 & \frac{q}{2} \lambda \langle k \rangle -
  \lambda -1 - w & 1 \\ 0 & 0 & 0 & 2 \lambda & -2 \end{array}
\right),
\end{eqnarray*}
and
\begin{eqnarray*}
{\bf I} \equiv \left( \begin{array}{ccccc} I_A & I_AS_A & I_AI_A &
  I_AS_B & I_AI_B \end{array} \right).
\end{eqnarray*}
At the critical point, the determinant of $\bf A$ must be zero, from
where
\begin{equation}
 \left( \lambda \langle k \rangle -1 \right) \left[ q \lambda \langle
   k \rangle - 2(1 + w) \right] - q \lambda^2 \langle k \rangle^2 = 0.
\label{eigenvalue-eq}
\end{equation}
Finally, solving for $\omega$ we obtain the critical rewiring rate
\begin{equation}
\omega_{c,q}^{\lambda} = \frac{(1+q/2) \lambda \langle k \rangle
  -1}{1- \lambda \langle k \rangle}.
\label{wcqlambda}
\end{equation}
All real parts of the eigenvalues of the Jacobian matrix ${\bf A}$
must be strictly negative  for the healthy state solution ${\bf I=0}$
to be stable.  Then, a sufficient condition for the instability is the
existence of a positive real part of one of the five eigenvalues.
This happens in two regions $R_1$ and $R_2$ of the $\omega-\lambda$
phase space that fulfill the conditions
\begin{eqnarray}
\lambda > \lambda_c ~~ &\mbox{and}& ~~ \omega \ge 0 ~~ \mbox{(region
  $R_1$)}
\label{cond1}
\\ \mbox{or} \nonumber \\ \lambda_{c,q} < \lambda < \lambda_c ~~
&\mbox{and}& ~~ 0< \omega < \omega_{c,q}^{\lambda} ~~ \mbox{(region
  $R_2$).~~~~~}
\label{cond2}
\end{eqnarray}
Regions $R_1$ and $R_2$ make up the endemic phase [see
  Fig.~\ref{fig:wc-lambda}(b)].  According to Eq.~(\ref{cond1}), when
$\lambda$ is above the epidemic threshold of the single network
$\lambda_c$ the epidemics propagates on the coupled system,
independently of the value of the rewiring.  This is in agreement with
the fact that in region $R_1$ the critical rewiring
$\omega_{c,q}^{\lambda}$ from Eq.~(\ref{wcqlambda}) is negative for
any value of the parameter's topology.  Since $\omega$ is meaningful
only for positive values there is no critical rewiring able to stop
the propagation of the epidemics.  At the single layer threshold
$\lambda_c$ the critical rate $\omega_{c,q}^{\lambda}$ diverges, and
it is above the region $R_2$ that for finite values of $\omega >
\omega_{c,q}^{\lambda}$ the two-network system becomes effectively
decoupled, and the epidemics cannot propagate.  Thus, a high enough
rewiring rate can induce an effective rescue of weak endemic states.

In Fig.~\ref{fig:wc-lambda}(a) we plot $\omega_{c,q}^{\lambda}$ vs
$\lambda$ from Eq.~(\ref{wcqlambda}) (solid lines).  The agreement
with the integration of Eqs.~(\ref{dIdtpa}) (dots) is perfect.
Figure~\ref{fig:wc-lambda}(b) shows the comparison with simulations of
the SIS dynamics on two ER networks with $N=10^4$ nodes, average
degree $\left< k \right> = 10$, and $q=1.0$ for the coupling. To
estimate the critical rewiring we performed spreading experiments.
They consist on running the dynamics from an initial configuration
with a few infected neighboring nodes in layers A and B, and
calculating the survival probability $S(t)$ of the run for various
values of $\lambda$, i e., the probability that a given run did not
reach the healthy state up to time $t$.  This probability shows an
algebraic decay $S \sim t^{-1}$ at the critical point.  We observe
that the critical rewiring rate obtained from the simulations (filled
circles) coincides with the one predicted by the system of
Eqs.~(\ref{dIdtpa}) (solid line), with a divergence at $1/ \left< k
\right> = 0.1$.

From the eigenvalue Eq.~(\ref{eigenvalue-eq}), we can also obtain the
critical infection rate as a function of $w$
\begin{equation}
\lambda_{c,q}^w = \frac{1}{\left( 1+q_w/2 \right) \langle k \rangle},
\label{lambda-cqw}
\end{equation}
where
\begin{equation}
q_w \equiv \frac{q}{1+w},
\label{qw}
\end{equation}
which can be interpreted as an \emph{effective coupling} when the
rewiring is present.  This is so because Eq.~(\ref{lambda-cqw}) has
the same structural form as the expression Eq.~(\ref{lambda_cq}) for
the critical infection rate in a static two-layer network with
coupling $q_w$. This result indicates that for a fixed
interconnectivity $q$, the critical infection rate grows with $\omega$
from $\lambda_{c,q}$ (for $\omega=0$) to $\lambda_c$ (for $\omega \to
\infty$) [see Fig.~(\ref{lambdacq})]. As we see, the effect of
rewiring is then to reduce the coupling between the two layers for the
transmission of the disease, even though the ``structural coupling''
$q$ is not affected by $w$.  Therefore, we can think the two-layered
system with \emph{dynamic} interconnections rewired at rate $w$ as a
system with \emph{static} interconnections but with an effective
reduced coupling $q_w$.  An important consequence of Eq.~(\ref{qw}) is
the fact that $q_w \to 0$ in the infinitely fast rewiring limit $w \to
\infty$, and thus the endemic phase region $R_2$ disappears.  This
confirms our intuitive idea that when the rewiring is extremely large
there are no $IS$ links at the interface and, therefore, layers are
effectively decoupled and behave like isolated networks.

\section{Strong endemic states}
\label{strong}
We explore in this section how the rewiring affects the epidemics when
infection rates are above the single layer threshold $\lambda_c$.  In
this case, the endemic state can be sustained in each network
separately, but we shall see that when the layers are interconnected
the prevalence depends on the coupling $q$ and the rewiring rate
$\omega$.  We first make the analysis using the rate equations, which
corresponds to the behavior on infinite large systems.  Then, we study
the case of finite systems, where new phenomenology appears due to the
extinction of the epidemics by finite size fluctuations.

\subsection{Thermodynamic limit}
\label{infinite}

\begin{figure}[t]
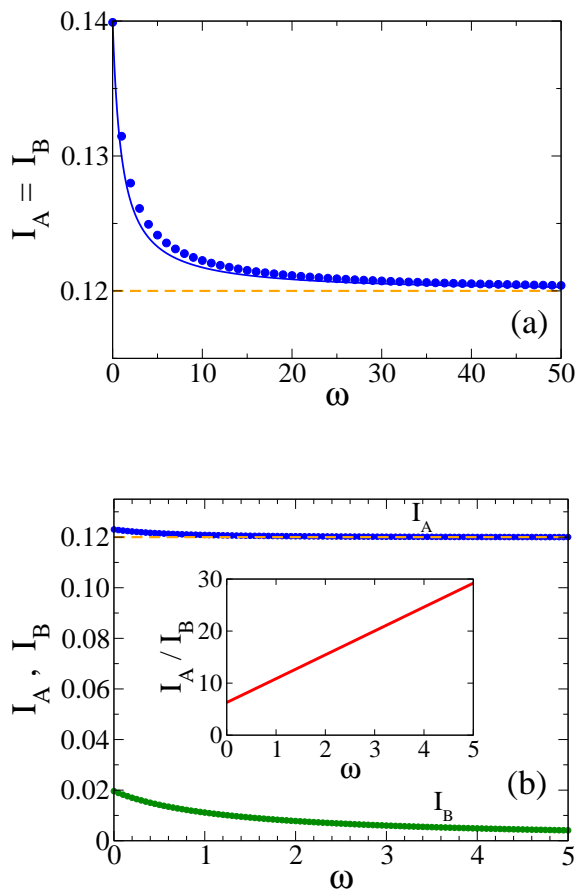

\begin{tabular}{l}
\includegraphics[width=7.5cm]{Fig4a.eps} \\\\\\\\
\includegraphics[width=7.5cm]{Fig4b.eps}
\end{tabular}
\caption{Prevalence $I_A$ and $I_B$ in networks A and B, respectively,
  as a function of the rewiring rate $\omega$.  In both graphs we used
  $\left<k\right>=10$ and $q = 0.5$.  (a)  The infectivities are
  $\lambda_A=\lambda_B=\lambda_{AB}=\lambda_{BA}=\lambda=0.115$ above
  the single layer critical rate $\lambda_c=0.1$.  The dashed line
  corresponds to the prevalence in network A or B when isolated.  (b)
  The infectivities are $\lambda_A=\lambda_{AB}=0.115$ and
  $\lambda_B=\lambda_{BA}=0.085$, above and below $\lambda_c$,
  respectively.  The prevalence in layer B (lower line) is zero when
  it is isolated, while the prevalence in layer A (upper line)
  approaches its single layer value $0.12$.  Inset: ratio between
  prevalence in networks A and B as a function of $\omega$.}
\label{fig:1}
\end{figure}

In Fig.~\ref{fig:1} we show the prevalence levels $I_A$ and $I_B$ as a
function of the rewiring rate $\omega$, obtained from the numerical
integration of equations~(\ref{dIdtpa}), using $\left<k\right>=10$ and
$\left< k_{AB} \right>=0.25$.  Fig.~\ref{fig:1}(a) corresponds to
infection rates
$\lambda_A=\lambda_B=\lambda_{AB}=\lambda_{BA}=\lambda=0.115$ above
the epidemic threshold $\lambda_c=1/\left< k \right> =0.1$, thus each
separate network is in the endemic state.  As expected, $I_A$ and
$I_B$ reach the same stationary value.  We observe that the prevalence
decreases monotonically as $\omega$ increases, and approaches the
prevalence in the isolated networks (dashed horizontal line) as
$\omega$ becomes very large.  This behavior is reminiscent of what we
found in the last section, that is, both networks become effectively
decoupled as $\omega$ goes to infinity, leading to stationary values
$I_A$ and $I_B$ corresponding to single isolated networks.  This
phenomenon can also be captured from Eqs.~(\ref{dIdtpa}), by
considering the limiting case $\omega \to \infty$.  In this limit, the
dominant term on the right hand side of Eq.~(\ref{dIASBdt}) is
$-\omega I_AS_B$, which gives the exponential time decay $I_AS_B(t) =
(I_AS_B)_0 \, e^{-\omega t}$, with $(I_AS_B)_0$ the initial density of
$I_AS_B$ pairs.  That is, the stationary density of $I_AS_B$ pairs is
zero.  Following the same argument we obtain from Eq.~(\ref{dSAIBdt})
that the stationary density of $S_AI_B$ pairs is also zero.  Then,
using these two stationary values in Eq.~(\ref{dIAIBdt}) we get that
$I_AI_B$ is zero as well, due to the combined effects of recovery and
rewiring.  Finally, from the conservation relation
Eq.~(\ref{conservAB}) we can see that $S_AS_B=\left< k_{AB} \right>$.
In other words, in the infinite rewiring limit the system quickly
evolves to a stationary state characterized by the absence of
interlinks connecting infected nodes, and thus all interlinks are
between susceptible neighbors only.  As the disease is only
transmitted through infected nodes, there is no possible spreading
between networks, which behave effectively as independent on the
spreading on its partner network.  The complete decoupling leads to a
set of three equations corresponding to the evolution of the epidemics
on a single network, by setting $I_AS_B=S_AI_B=I_AI_B=0$ in
Eqs.~(\ref{dIAdt}), (\ref{dIASAdt}) and (\ref{dIAIAdt}), whose
stationary solution $I_A$ is the dashed line in Fig.~\ref{fig:1}(a).

As was shown in~\cite{Saumell:2012}, it is not possible to find a
mixed phase in the coupled static system, that is, a phase that
consists on an endemic state in one layer and a healthy state in the
other layer.  This is so because when the disease is able to propagate
on one layer, then it always propagates on the entire system.  This
phenomenon is also observed even if the interlinks are rewired, as we
see in Fig.~\ref{fig:1}(b), where we show results from
Eq.~(\ref{dIdtpa}) with infection rates
$\lambda_A=\lambda_{AB}=0.115>\lambda_c=0.1$ and
$\lambda_B=\lambda_{BA}=0.085<\lambda_c=0.1$.  Under these asymmetric
conditions, the disease is able to spread in layer A but it dies off
in layer B, when the networks are disconnected.  Again, finite
rewiring rates reduce the prevalence in each layer but cannot induce a
complete breaking of the coupling, which only happens in the $\omega
\to \infty$ limit.

\subsection{Finite size effects}

In the previous sections we studied the prevalence of the disease in
both networks under different scenarios, and how this prevalence is
affected by the rewiring of internetwork links, as compared to the
case of static interconnections.  The analysis was done by means of a
system of equations for the densities of nodes and pairs, which is a
good mean-field description of the disease dynamics in infinite large
networks, where finite size fluctuations are neglected.  This allows
to study various phenomena that are independent on the system size for
large enough networks, such as the transition between the endemic and
healthy phase when $\lambda$ and $\omega$ are varied.  These equations
predict that the disease in the endemic phase last for infinite long
times.  However, in finite networks, fluctuations in the density of
infected nodes in a given realization eventually drive the system to
an absorbing configuration, where all nodes are susceptible.  This
brings new interesting phenomena that are intrinsic of finite systems,
and that we explore in this section.

\begin{figure}[t]
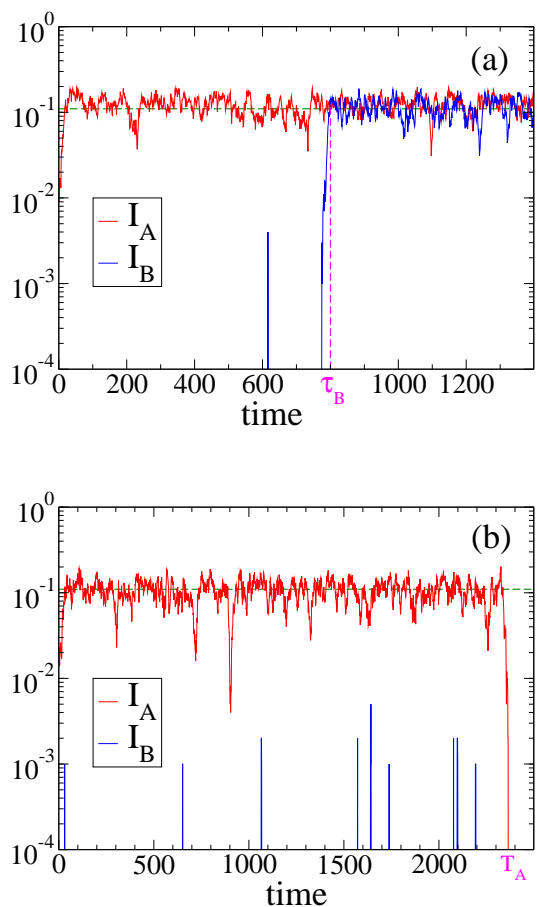

\begin{tabular}{l}
\includegraphics[width=7.0cm]{Fig5a.eps} \\\\\\ 
\includegraphics[width=7.0cm]{Fig5b.eps}
\end{tabular}
\caption{Time evolution of the density of infected nodes $I_A$ and
  $I_B$ in networks A and B, respectively, in two distinct
  realizations of the dynamics (a) and (b), on two coupled ER networks
  of $N=10^3$ nodes each, coupling $q=0.005$ and interlink rewiring
  $\omega=50$ and
  $\lambda_A=\lambda_{AB}=\lambda_B=\lambda_{BA}=\lambda=0.115 >
  \lambda_c=0.1$.  In realization (a) there is an outbreak in both
  layers, while in realization (b) the outbreak occurs only in layer
  A.}
\label{fig:one-rea}
\end{figure}

We are interested in studying whether the rewiring is able to stop the
spreading of the disease from network A to network B, when
fluctuations are taken into account.  As we showed in section
\ref{dynamic}, the rewiring reduces the effective connectivity between
both networks and, as a consequence, it can bring the system from the
endemic to the healthy phase when it overcomes a critical rewiring
$\omega_{c,q}^{\lambda}$.  As explained in section \ref{infinite},
this is only possible when the infection rates of both networks
$\lambda_A$ and $\lambda_B$ are below the critical infection rate of
an isolated network $\lambda_c$, otherwise the disease spreads over
both networks, independently on the initial condition and the rewiring
rate.  In other words, when either $\lambda_A>\lambda_c$ or
$\lambda_B>\lambda_c$, if the disease spreads on A then it spreads on
B with probability equal to unity.  Only in the $\omega \to \infty$
limit spreading is stopped, because networks are effectively
disconnected.

However, we shall see that this scenario is no longer valid in finite
networks, where in a single realization of the dynamics with finite
$\omega$ (including $\omega=0$), the spreading  can take place in one
network only.  Indeed, Figs.~\ref{fig:one-rea}(a) and (b) show two
distinct realizations with the same parameters and initial condition,
but with very different outcomes.  Initially, only one random node in
network A and its neighbors are infected.    We investigate whether
this initial small focus of infection propagates through network A,
and is able to invade network B.  In Fig.~\ref{fig:one-rea}(a), the
disease in network A quickly spreads  until it reaches a stationary
state in which $I_A$ fluctuates around a given value $I^s \simeq 0.11$
(horizontal dashed line).  We refer to this state where a large
fraction of the population gets infected as an \emph{outbreak}.  In
network B the outbreak happens at a much longer time $\tau_B \simeq
750$, but eventually the disease spreads on both networks, as it
happens in infinite systems.  However, in fig.~\ref{fig:one-rea}(b)
there is also a quick outbreak in network A but the outbreak in
network B never happens, because $I_A$ becomes zero by a large
fluctuation at time $T_A$, driving the entire system to a halt.  That
is,   the epidemics on A goes extinct before the disease can spread on
network B.  We observe that in various occasions a small fraction of
nodes in network B -probably those with connections to network A- get
infected, rising $I_B$ to values larger than zero that quickly die
out, producing a ``spike-like'' pattern in $I_B$.  Nevertheless, it
seems that $I_B$ never reaches a level that is large enough to
initiate a cascade of infections that leads to an outbreak.

In the next section we study the statistics of outbreak times $\tau_B$
and their associated probabilities.  We shall see that the rewiring
amplifies finite size effects,  inducing a delay in the outbreak of
network B that increases with $\omega$, so that the outbreak may
remain confined in network A for arbitrary long times.  We also show
that the there is a finite probability that the disease does not
spread on network B.

\subsubsection{Outbreak probabilities and times}

We ran MC simulations of the dynamics on two coupled ER networks with
$N$ nodes each, mean degrees $\langle k \rangle = 10$, coupling
$q=0.005$ and infection rates
$\lambda_A=\lambda_{AB}=\lambda_B=\lambda_{BA}=\lambda=0.115$.
Because $\lambda$ is  above the critical value $\lambda_c=0.1$,
typically an initial seed in network A spreads over a large fraction
of the system.  When an outbreak happens in A, then the disease can
pass to network B through interlinks, and may cause an outbreak in B
as well.  As we mentioned earlier this is not always the case, because
sometimes single infected nodes are not able to initiate an avalanche
of massive infections and the disease dies out.  We want to estimate
the probability $P_{B|A}$ that there is an outbreak in network B given
that there was an outbreak in network A.  This conditional probability
measures the fraction of times that finite size effects amplified by
rewiring are not able to stop the ``transmission'' of an outbreak from
A to B.  To calculate the outbreak statistics in B conditioned to
outbreaks in A, we start the runs from a situation where there is
already an outbreak in A, by randomly infecting a fraction $I_A=0.12$
of nodes in A, above the prevalence level $I^s=0.11$.  Starting from a
seed in A is not computationally efficient because there is a chance
that the disease out in A before an outbreak occurs and, therefore, an
outbreak in B neither takes place.

\begin{figure}[t]
  \begin{tabular}{l}
    \includegraphics[width=7.0cm]{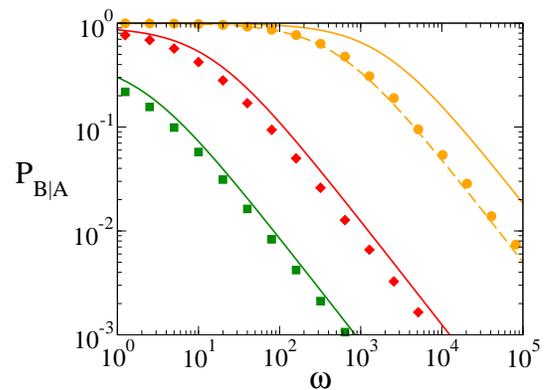}
  \end{tabular}
  \caption{Probability $P_{B|A}$ of an outbreak in layer B after an
    outbreak happened in layer A of a coupled two-layer system, as a
    function of the interlink rewiring rate $\omega$.  Each layer is
    an ER network with mean degree $\left< k \right>=10$.  The
    coupling is $q=0.005$ and the infection rates are $\lambda=0.115$.
    Symbols represent MC simulation results for different system
    sizes: $N=250$ (squares), $N=500$ (diamonds) and $N=1000$
    (circles).  Solid lines are the analytic approximations from
    Eq.~(\ref{PBA-analy}), while the dashed line is the approximation
    Eq.~(\ref{PBA}) using the values $\mu=6.49 \times 10^{-5}$ and
    $\beta=0.0328/(1+\omega)$ calculated in Appendices \ref{muapen}
    and \ref{betaapen}, respectively.}
  \label{fig:PBA-w}
\end{figure}

In Fig.~\ref{fig:PBA-w} we plot the outbreak probability in network B,
$P_{B|A}$, as a function of the rewiring rate $\omega$, obtained from
MC simulations for various system sizes (solid symbols).  We observe
that $P_{B|A}$ decays very slowly when $\omega$ increases, from a
value $P_{B|A}(\omega=0)$ that approaches $1$ as $N$ becomes large
($P_{B|A}(\omega=0) = 0.881$ and $0.998$ for $N=500$ and $1000$,
respectively).  That  is, for a static or very slowly varying
internetwork connectivity, once the disease spreads on A then it also
spreads on B with high probability. But as $\omega$ increases, $I_A
S_B$ links last for shorter  times, decreasing the probability of
infection of B-nodes and the subsequent outbreak in network B.  For
$\omega > 5 \times 10^4$ and $N=1000$, the  outbreak in B is very
unlikely (smaller than $1\%$) , meaning that in most realizations the
spreading on B does not happen.  Therefore, the rewiring in finite
networks proves to be very efficient in preventing the transmisibility
between networks.  In the next section we show that $P_{B|A}$ decays
as $\omega^{-1}$  for $\omega \gg 1$ (solid lines in
Fig.~\ref{fig:PBA-w}), thus $P_{B|A}$ becomes zero only in the
infinite large rewiring limit $\omega \to \infty$.

\begin{figure}[t]
\begin{tabular}{l}
\includegraphics[width=7.0cm]{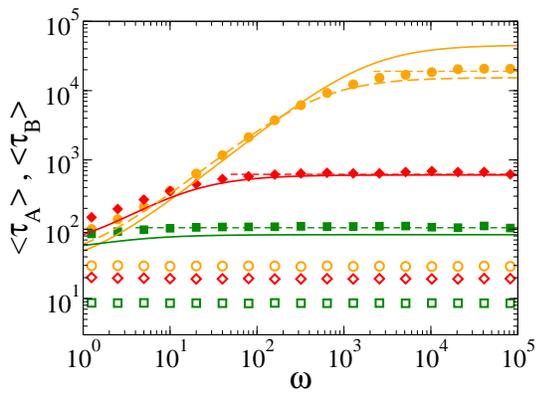}
\end{tabular}
\caption{Mean outbreak times $\left< \tau_A \right>$ (open symbols)
  and  $\left< \tau_B \right>$ (filled symbols) in networks A and B,
  respectively, vs the rewiring rate $\omega$, for a two-layer system
  with the same parameters and system sizes as in
  Fig.~\ref{fig:PBA-w}.  Solid curves are the analytic approximations
  from Eq.~(\ref{tauB-analy}), while the dashed curve is the
  approximation Eq.~(\ref{tauB}) with the values $\mu=6.49 \times
  10^{-5}$ and $\beta=0.0328/(1+\omega)$ calculated in Appendices
  \ref{muapen} and \ref{betaapen}, respectively.  Horizontal dashed
  lines correspond  to numerical results of the mean extinction times
  in network A.}
\label{fig:TBA-w}
\end{figure}

In Fig.~\ref{fig:TBA-w} we show mean outbreak times  $\langle \tau_A
\rangle$ and $\langle \tau_B \rangle$ in networks A and B,
respectively, over many independent realizations as a function of
$\omega$.  The mean time $\langle \tau_A \rangle$  ($\langle \tau_B
\rangle$) is an average only over realizations in which network A
(network B) undergoes an outbreak.  Outbreak times in A were
calculated using a seed in A as initial condition.  We observe in
Fig.~\ref{fig:TBA-w} that $\langle \tau_A \rangle$ is very short and
independent on $\omega$, while $\langle \tau_B \rangle$ increases
monotonically with $\omega$ and reaches a constant value as $\omega
\to \infty$.  The saturation level $\langle T_A \rangle$ corresponds
to the mean extinction time in network A when isolated (see
Eq.~(\ref{Tup}) in Appendix \ref{muapen} for the scaling of $\langle
T_A \rangle$  with the model's parameters).  This result
is a direct consequence of the fact that the outbreak in B is limited
by the extinction in A.  Actually, in the very large rewiring limit
$\omega \gg 1$ outbreaks in B are very unlikely, but when a B-outbreak
does take place it usually happens just before the extinction in
network A, so that $\tau_B \lesssim T_A$ in each realization, and
therefore $\langle \tau_B \rangle \lesssim \langle T_A \rangle$.

In the next section we develop an approach based on a reduced model
specially designed to account for the stochastic dynamics of the
coupled system of two networks, which allows to obtain analytic
expressions for the outbreak probabilities and times.

\subsubsection{Three-state symbolic model}

In order to gain an insight about the statistics of outbreak times and
their probabilities we need to consider the fluctuations associated
with the system size.  One way to incorporate the stochastic dynamics
is to derive a system of equations as the one in section \ref{math},
but with some additional noise terms corresponding to finite-size
fluctuations.  This procedure is in general very tedious and hard to
carry out because of the large number of processes and equations
involved and, besides, only numerical solutions can be obtained.  That
is why we develop here a much simpler approach, based on a synthetic
model specially useful to investigate the outbreak probabilities and
times.  The approach consists on viewing the system and its dynamics
at a coarse-grained level.  We represent each network as a simple unit
(node), and the state of each network as a binary variable, which is a
symbolic representation of the two possible prevalence levels:
outbreak when the density of infected nodes $I$ fluctuates around
$I^s$ (active state), and no-outbreak when  $I$ is zero or close to
zero (inactive state).  This approach is an oversimplification of the
complex dynamics of the system, but is able to reproduce rather well
the collective properties of extinction and outbreak processes.

\begin{figure}[t]
\begin{tabular}{ll}
\includegraphics[width=4.0cm]{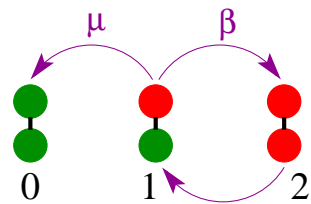}
\end{tabular}
\caption{Schematic representation of the synthetic model.  A red node
  represents an active network, while a green node denotes an inactive
  network.  $\mu$ and $\beta$ are the effective transition rates
  between the three possible states of the two-network system.}
\label{fig:AB}
\end{figure}

In Fig.~\ref{fig:AB} we show the possible states of the two-node
system and their associated transitions.  Red and green nodes
represent, respectively, active and inactive networks.  States 0, 1
and 2 denote the situations in which both networks are inactive, only
one network is active, and both networks are active, respectively.
These states represent the total activity of the system, so that the
two cases A-active/B-inactive and A-inactive/B-active are grouped into
a single state 1.  An active node becomes inactive at rate $\mu$, and
activates its neighboring node at rate $\beta$.  The dynamics of this
synthetic model is analogous to the one of the SIS model, but with
macroscopic transition rates $\mu$ and $\beta$ that are
\emph{effective rates} that depend on the model's parameters: the
network size $N$, the infection rate $\lambda$, the mean connectivity
$\langle k \rangle$, the interconnectivity $q$, and the rewiring rate
$\omega$.  This model tries to capture the time evolution of the
outbreak/extinction behavior of the coupled system of two networks.
For instance, if at a given time network A is active and  network B is
inactive (state $1$), we assume that network A  triggers an outbreak
in network B at constant rate $\beta$ ($1 \to 2$ transition), as in
Fig.~\ref{fig:one-rea}(a), and that the disease in network A dies out
at constant rate $\mu$ ($1 \to 0$ transition), as  in
Fig.~\ref{fig:one-rea}(b).

Within this reduced model, the dynamics of transitions between states
follows a multiple Poisson process with rates $\mu$ and $\beta$.
Therefore, the outbreak probability in network B given an outbreak in
A can be estimated as
\begin{equation}
P_{B|A} \simeq P_{12} = \frac{\beta}{\mu+\beta},
\label{PBA}
\end{equation}
and the mean outbreak time in network B as
\begin{eqnarray}
\left< \tau_B \right> \simeq \left< \tau_{12} \right> =
\frac{1}{\mu+\beta}.
\label{tauB}
\end{eqnarray}
Here $\left< \tau_{12} \right>$ is the mean time of the $1 \to 2$
transition normalized by $P_{12}$, that is, over those realizations
that made the transition to the outbreak.  To check the validity of
Eqs.~(\ref{PBA}) and (\ref{tauB}) we run MC simulations to calculate
the rates $\mu$ and $\beta$ for networks of size $N=10^3$ (see
Appendices~\ref{muapen} and \ref{betaapen} for details).  Dashed lines
in Figs.~\ref{fig:PBA-w} and \ref{fig:TBA-w} correspond to
Eqs.~(\ref{PBA}) and (\ref{tauB}) using the numerical values of $\mu$
and $\beta$.  We observe that the agreement is good, showing that the
symbolic model describes quite well the stochastic macroscopic
dynamics of the system.

In Appendices~\ref{muapen} and \ref{betaapen} we also develop an
analytical approach to obtain approximate expressions for $\mu$
[Eq.~(\ref{mu})] and $\beta$ [Eq.~(\ref{beta})].  Plugging
Eqs.~(\ref{mu}) and (\ref{beta}) into Eqs.~(\ref{PBA}) and
(\ref{tauB}) we obtain the following expressions for the outbreak in
$B$,
\begin{eqnarray}
\label{PBA-analy}
P_{B|A} &\simeq& \left[ 1+\frac{\mathcal A \, e^{-\mathcal B N}
    \,(1+\omega)}{\sqrt{N} \, q} \right]^{-1}, \\ \nonumber \\
\label{tauB-analy}
\left< \tau_B \right>  &\simeq& \left[ \mathcal C \, \sqrt{N} \, e^{-
    \mathcal B N} + \frac{\mathcal D \, q \, N}{1+\omega}
  \right]^{-1},
\end{eqnarray}
where $\mathcal{A = C/D}$, and the coefficients $\mathcal{B, C}$ and
$\mathcal D$ are given in Eqs.~(\ref{B-coef}), (\ref{C-coef}) and
(\ref{D-coef}) of the Appendices.  Solid curves in
Figs.~\ref{fig:PBA-w} and \ref{fig:TBA-w} correspond  to
Eqs.~(\ref{PBA-analy}) and (\ref{tauB-analy}).  Even though we observe
a discrepancy with the numerical values (symbols) that increases with
the system size, Eqs.~(\ref{PBA-analy}) and (\ref{tauB-analy})
correctly capture the qualitative behavior of $P_{B|A}$ and $\left<
\tau_B \right>$ with the model's parameters.

\begin{figure}[t]
\includegraphics[width=7.0cm]{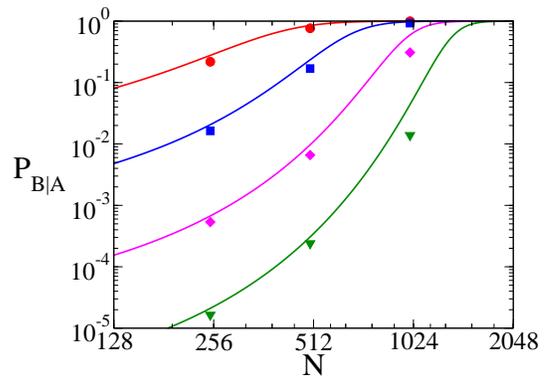}
\caption{Outbreak probability in network B vs system size $N$ for four
  different values of the rewiring rate $\omega=1.25$ (circles),
  $\omega=40$ (squares), $\omega=1280$ (diamonds) and $\omega=40960$
  (down triangles).  Solid curves are the analytic approximations from
  Eq.~(\ref{PBA-analy}).}
\label{fig:PBA-N}
\end{figure}

We observe from Eq.~(\ref{PBA-analy}) and Fig.~\ref{fig:PBA-w} that
for finite and large $N$, $P_{B|A} \lesssim 1$ in a static network
$\omega=0$, but it rapidly decreases with the rewiring as
$(1+\omega)^{-1}$.  Also, $P_{B|A} \to 1$ in the thermodynamic limit,
for all finite $\omega$ (see Fig.~\ref{fig:PBA-N}).  This is in
agreement with the mentioned fact that for infinite large systems in
the strong endemic state, an outbreak in A always implies an outbreak
in B.  In this case, as layer A stays for ever in the endemic state,
even if the transmission rate to layer B is reduced to very small
values by decreasing $q$ or increasing $\omega$, the disease always
ends up invading and breaking up on B.  Therefore, we conclude that
the rewiring has no effect on the outbreak probability for infinite
large networks, but as long as the networks are finite, its effect is
amplified as $N$ decreases.  From Eq.~(\ref{tauB-analy}) and
Fig.~\ref{fig:TBA-w} we see that for $\omega \to 0$ mean outbreak
times in both networks are similar, while in the opposite limit
$\omega \to \infty$, the mean outbreak time in layer B $\left< \tau_B
\right>$ approaches $1/\mu$, which is similar to the mean extinction
time in network A $\left< T_A \right>$ (horizontal dashed lines).

\section{Summary and Conclusions}
\label{conclusions}

We studied a model for disease spreading on two layers of networks
coupled through dynamic interconnections.  Links connecting infected
nodes in different layers are rewired at rate $\omega$ to avoid
interlayer contacts between infected and susceptible nodes, so that
the interconnecting network coevolves dynamically with the dynamics
of the states of the nodes in the two layers. We developed an
analytical approach based on the pair approximation to explore the
system's dynamics.  This approach reveals that the effect of the
rewiring is to reduce the effective coupling between the two layers
for the disease transmission, even though the structural coupling is
not affected by $\omega$.  We find two different mechanisms by which
rewiring can stop the propagation of the disease, effectively rescuing
endemic states of the global statically interconnected network. Each
of these mechanisms is efficient in one of the two scenarios
considered of weak or strong endemic states.

In weak endemic states in which the infection rate is below the
epidemic threshold of a single layer, the two-layer system with
dynamic interconnections can be thought as a system with static
interconnections, but with an effective coupling that monotonically
decreases with $\omega$.  In this case, a high enough rewiring can
reduce to zero the disease prevalence found with a static
interconnection of layers.  This is described in statistical terms by
a transition from the endemic to the healthy phase as $\omega$
overcomes a given threshold, which increases with the coupling and the
networks' connectivity.

In strong endemic states in which the infection rate is above the
epidemic threshold of a single layer, we studied spreading between
layers, starting from a seed of infected nodes in one layer and
spreading to the second layer. For a finite size system there is a
probability that the infection dies out in the first layer before
propagating to the second layer. This probability is practically
negligible for a static interconnection, but finite size effects are
amplified by interlink rewiring, leading to a finite probability that
the epidemics does not propagate to the second layer. The
complementary probability that the disease spreads from the infected
layer to the susceptible layer in a finite system decreases
monotonically with $\omega$, and approaches zero as $\omega \to
\infty$.

In summary, our results show that the adaptive rewiring of interlayer
links is an effective strategy to control spreading across communities
of individuals, either by the existence of a critical rewiring rate
that effectively decouples the two communities, or by the
amplification of finite size effects that prevent propagation of the
disease from one community to the other.

\section*{ACKNOWLEDGMENTS}
We acknowledge financial support from EC project LASAGNE
FP7-2012-STREP-318132. MAS acknowledges support from the James
S. McDonnell Foundation 21st Century Science Initiative in Studying
Complex Systems Scholar Award, MINECO Project FIS2013-47282-C02-01,
and the Generalitat de Catalunya grant 2014SGR608. MSM also
acknowledges support from MINECO (Spain) and FEDER under Project
FIS2012-30634 (INTENSE@COSYP).  FV also acknowledges support from
CONICET (Argentina).

\appendix
\section{Pair approximation approach}
\label{pair}

The time evolution of the model can be approximately described by the
following set of ODEs for nodes and links, which account for state
correlations between nearest-neighbors in the networks (first-order
correlations):
\begin{subequations}
\label{dIdtsys}
\begin{eqnarray}
\frac{dI_A}{dt} &=& -I_A + \lambda_A I_AS_A + \lambda_{BA} S_AI_B,
\\ \frac{dI_AS_A}{dt} &=& 2 I_AI_A - (1+\lambda_A)I_AS_A - 2 \lambda_A
I_AS_AI_A \\ &-& \lambda_{BA} I_AS_AI_B + \lambda_A S_AS_AI_A +
\lambda_{BA} S_AS_AI_B, \nonumber \\ \frac{dI_AI_A}{dt}&=&- 2 I_AI_A +
\lambda_A I_AS_A + 2\lambda_A I_AS_AI_A  \nonumber \\ &+& \lambda_{BA}
I_AS_AI_B, \\ \frac{dI_AS_B}{dt}&=& I_AI_B - (1 + \omega) I_AS_B -
\lambda_{AB} I_AS_B \nonumber \\ &+& \lambda_A I_AS_AS_B  - \lambda_B
I_AS_BI_B - 2 \lambda_{AB} I_AS_BI_A \nonumber \\ &+&  \lambda_{BA}
I_BS_AS_B, \\ \frac{dS_AI_B}{dt}&=& I_AI_B - (1 + \omega) S_AI_B -
\lambda_{BA} S_AI_B \nonumber \\ &+& \lambda_B S_AS_BI_B  - \lambda_A
I_AS_AI_B - 2 \lambda_{BA} I_BS_AI_B \nonumber \\ &+&  \lambda_{AB}
S_AS_BI_A, \\ \frac{dI_AI_B}{dt}&=&- 2  I_AI_B + \lambda_{AB} I_AS_B +
\lambda_{BA} S_AI_B \nonumber \\ &+& \lambda_A I_AS_AI_B  + \lambda_B
I_AS_BI_B  + 2 \lambda_{AB} I_AS_BI_A \nonumber \\ &+& 2 \lambda_{BA}
I_BS_AI_B.
\end{eqnarray}
\end{subequations}
We have omitted here the rate equations for $I_B$, $I_BS_B$ and
$I_BI_B$ because they have the same form as the equations for $I_A$,
$I_AS_A$ and $I_AI_A$, respectively, after interchanging sub-indices
$A$ and $B$.  Also, densities $S_A$, $S_B$, $S_AS_A$, $S_BS_B$ and
$S_AS_B$ can be obtained from the conservation relations
Eq.~(\ref{conserv}).  We note that the system of
equations~(\ref{dIdtsys}) expresses the evolution of the densities
that contain at least one infected node. We shall see that this is
particularly convenient when we analyze the stability of the healthy
phase.

As we can see, the rate equations for pairs depend on the densities of
triplets, for which we use a notation analogous to the ones used for
nodes and links.  For instance, $I_AS_BI_A$ represents the density per
node of triples consisting of an infected node in A connected to a
susceptible node in B that is in turn connected to another infected
node in A.  To close the system of equations ~(\ref{dIdtsys}) we make
use of the pair approximation, which assumes that the links of
different types are homogeneously distributed over the networks.  This
allows to decompose triples in terms of nodes and pairs (see for
instance \cite{Keeling:1999,Gross:2006,Demirel:2014,Shai:2013}), closing the
system at the level of pairs.  The triplets of Eq.~(\ref{dIdtsys}) can
be expressed as
\begin{eqnarray}
\label{triplets}
I_AS_AI_A &=& \frac{I_AS_A \cdot I_AS_A}{2 S_A} \, ; ~~ I_AS_AI_B =
\frac{I_AS_A \cdot S_AI_B}{S_A}, \nonumber \\ I_AS_BI_A &=&
\frac{I_AS_B \cdot I_AS_B}{2 S_B} \, ; ~~~ I_AS_BI_B = \frac{I_AS_B
  \cdot I_BS_B}{S_B}, \nonumber \\ I_AS_AS_A &=& \frac{2 I_AS_A \cdot
  S_AS_A}{S_A} \, ; ~~ I_AS_AS_B = \frac{I_AS_A \cdot S_AS_B}{S_A},
\nonumber \\ I_AS_BS_A &=& \frac{I_AS_B \cdot S_AS_B}{S_B} \, ; ~~~
I_AS_BS_B = \frac{2 I_AS_B \cdot S_BS_B}{S_B}. \nonumber \\
\end{eqnarray}
Replacing the approximate expressions for the triplets from
Eq.~(\ref{triplets}) into the Eqs.~(\ref{dIdtsys}), we arrive to the
system of Eqs.~(\ref{dIdtpa}) of the main text.

\section{Prevalence in a symmetric two-layer system}
\label{prev}

In this section we derive an analytic expression for the stationary
density of infected nodes in a two-layer system with symmetric
conditions.  For the sake of simplicity, we first analyze the case of
a single islolated, and then adapt the results for the case of two
coupled networks.

Applying the pair approximation developed in Appendix \ref{pair} for
the case of a single network, for instance network A, we can describe
the evolution of the system by the set of equations
\begin{subequations}
\label{dIdtpasys}
\begin{eqnarray}
\label{dIAdt}
\frac{dI_A}{dt}&=& -I_A + \lambda I_AS_A, \\
\label{dIASAdt}
\frac{dI_AS_A}{dt} &=& 2I_AI_A - (1+\lambda) I_AS_A - \lambda
\frac{(I_AS_A)^2}{S_A} ~~~~~~\\ &+& 2 \lambda \frac{S_AS_A \cdot
  I_AS_A}{S_A}, \nonumber \\
\label{dIAIAdt}
\frac{dI_AI_A}{dt}&=&- 2 I_AI_A + \lambda I_AS_A  +  \lambda
\frac{(I_AS_A)^2}{S_A},
\end{eqnarray}
\end{subequations}
together with the conservation relations
\begin{subequations}
\label{conservAB2}
\begin{eqnarray}
\label{cons1}
1 &=& I_A+S_A, \\
\label{cons2}
\frac{\left<k\right>}{2} &=& S_AS_A+I_AI_A+I_AS_A,
\end{eqnarray}
\end{subequations}
where $I_A$, $I_AS_A$ and $I_AI_A$ are, respectively, the densities of
$I_A$ nodes, $I_AS_A$ links and $I_AI_A$ links.  Notice that these
equations can be derived from Eqs.~(\ref{dIdtpa}) by setting
$\lambda_{BA}=0$ (A isolated from B).  To find the stationary density
of infected nodes $I_A^s$ from Eqs.~(\ref{dIdtpasys}), we set the
derivatives to zero and express the stationary densities of links
$[S_AS_A]^s$ and $[I_AS_A]^s$ and $[I_AI_A]^s$ in terms of $I_A^s$,
using the relation Eq.~(\ref{cons1}).  We obtain
\begin{eqnarray*}
\left[ S_A S_A \right]^s &=& \frac{1-I_A^s}{2 \lambda} \\ \left[ I_A
  S_A \right]^s &=& \frac{I_A^s}{\lambda} \\ \left[ I_A I_A \right]^s
&=& \frac{I_A^s}{2} \left[1+\frac{I_A^s}{\lambda (1-I_A^s)} \right].
\end{eqnarray*}
Finally, plugging these expressions into Eq.~(\ref{cons2}) we arrive
to the following quadratic equation for $I_A^s$
\begin{equation}
\lambda (I_A^s)^2 - \lambda (1+ \left< k \right>) I_A^s + \lambda
\left< k \right> -1 = 0.
\label{Iseq}
\end{equation}
Only the solution of Eq.~(\ref{Iseq}) corresponding to the negative
branch has a physical meaning, which reads
\begin{equation}
I_A^s = \frac{1+\lambda_c}{2 \lambda_c} - \sqrt{
  \left(\frac{1-\lambda_c}{2 \lambda_c} \right)^2 + \frac{1}{\lambda}}
,
\label{Is}
\end{equation}
where $\lambda_c = \left< k \right>^{-1}$ is the critical infection
rate of the network.

In the symmetric and homogeneous case scenario of two identical
coupled networks A and B, with mean degree $\left< k \right>$,
infection rates
$\lambda_A=\lambda_{AB}=\lambda_{BA}=\lambda_B=\lambda$ and coupling
$q$, the static two-network system can be treated as a single network
with mean degree $\left< k \right> + \left< k_{AB}
\right>=(1+q/2)\left< k \right>$ and infection rate $\lambda$.
Therefore, using Eq.~(\ref{Is}), the stationary density of infected
nodes in the symmetric two-layer system is
\begin{equation}
I^s_{A,q} = \frac{1+\lambda_{c,q}}{2 \lambda_{c,q}} - \sqrt{
  \left(\frac{1-\lambda_{c,q}}{2 \lambda_{c,q}} \right)^2 +
  \frac{1}{\lambda}} ,
\label{Isq}
\end{equation}
where $\lambda_{c,q} = [( 1+\frac{q}{2})\left< k \right>]^{-1}$ is the
critical infection rate of the static coupled system.  For a dynamic
system with interlink rewiring $\omega$, we showed in section
\ref{dynamic} that the coupling becomes $q_{\omega} = q/(1+\omega)$,
and thus we expect the stationary density of infected nodes to behave
as
\begin{equation}
I^s_{A,q,\omega} = \frac{1+\lambda_{c,q}^{\omega}}{2
  \lambda_{c,q}^{\omega}} - \sqrt{
  \left(\frac{1-\lambda_{c,q}^{\omega}}{2 \lambda_{c,q}^{\omega}}
  \right)^2 + \frac{1}{\lambda}} ,
\label{Isw}
\end{equation}
with $\lambda_{c,q}^{\omega} = [(1+\frac{q_{\omega}}{2}) \left< k
  \right>]^{-1}$ the infection threshold.  Equation~(\ref{Isw}) agrees
reasonably well with the numerical integration of the system of
Eqs.~(\ref{dIdtpa}) for the two-network system [see
  Figs.~\ref{fig:I-w} and \ref{fig:1}(a)].  The agreement is very good
for $\omega$ close to $\omega_{c,q}^\lambda$, because the ansatz
relation Eq.~(\ref{qw}) that we used for the effective coupling is
only exact at the transition point (as we derived in section
\ref{dynamic}), that is, when $I_A=I_B=0$.  The agreement is also
perfect for $\omega=0$, because the solution Eq.~(\ref{Isq}) is exact,
while discrepancies arise between between $\omega=0$ and
$\omega_{c,q}^\lambda$

\section{Calculation of the extinction rate $\mu$}
\label{muapen}

To calculate the effective rate $\mu$ at which the epidemics
extinguishes on a single isolated network, we run MC simulations on an
ER network of $N=10^3$ nodes, mean degree $\left< k \right>=10$ and
infection rate $\lambda=0.115$, starting from an initial density
$I(0)=0.11$ of infected nodes that corresponds to the stationary
active state.  Then, we calculate the survival probability $S(t)$,
that is, the probability that the epidemics is not extincted up to
time $t$.  Working with survival probabilities rather than
first-passage probabilities is better because fluctuations  are
smaller.  Within the synthetic model, where we assume a constant decay
rate from the active to the inactive state, we expect an exponential
decay of the form $e^{-\mu t}$ for the survival probability.  In
simulations, this decay is observed after a very short initial
transient $\hat{t} \simeq 50$, during which $S$ is nearly constant.
Thus, we can approximate the shape of $S$ as
\begin{eqnarray}
 S(t) \simeq \left\{ \begin{array}{ll} 1 & \mbox{for}~~t \le \hat{t}
   \\ e^{-\mu(t-\hat{t})} & \mbox{for}~~t > \hat{t}.
\end{array} \right. \nonumber
\end{eqnarray}
The slope of the $S$ vs $t$ curve on a linear-log plot gives $\mu
\simeq 6.49 \times 10^{-5}$ (dashed curve in Fig.~\ref{fig:S1-t}).

An analytical expression for the upper bound of the average life time
$\left< T \right>$ of the SIS model on an arbitrary network was given
by Van Mieghem in \cite{Van-Mieghem:2013}.  It reads,
\begin{eqnarray}
\left< T \right> \simeq \frac{ \frac{\lambda}{\lambda_c} \, \sqrt{2
    \pi} \, \exp \Big\{ \left[ \ln \left( \frac{\lambda}{\lambda_c}
    \right) + \frac{\lambda_c}{\lambda} -1 \right] N \Big\} } {\left(
  \frac{\lambda}{\lambda_c} -1 \right)^2  \sqrt{N} },
\label{Tup}
\end{eqnarray}
where $N$ is the network size and $\lambda_c$ is the epidemic
threshold.  Plugging the expression $\lambda_c=1/\left< k \right>$ for
an ER network on Eq.~(\ref{Tup}), and given that $\mu \simeq 1/\left<
T \right>$, we arrive to the following expression for a lower bound of
the extinction rate
\begin{eqnarray}
\label{mu}
\mu &\simeq& \mathcal C \, \sqrt{N} \, e^{-\mathcal B N},
\end{eqnarray}
with
\begin{eqnarray}
\label{B-coef}
\mathcal B &\equiv& \ln \left( \lambda \left< k \right> \right) +
\frac{1}{\lambda \left< k \right>} - 1 ~~~~ \mbox{and}  \\
\label{C-coef}
\mathcal C &\equiv& \frac{ \left( \lambda \left< k \right> -1
  \right)^2} { \lambda \left< k \right> \sqrt{2 \pi} }.
\end{eqnarray}
For $N=10^3$, $\left< k \right>=10$ and $\lambda=0.115$ we get $\mu
\simeq 2.2 \times 10^{-5}$, which is approximately $1/3$ of the
numerical value $\mu \simeq 6.49 \times 10^{-5}$ obtained from
simulations.  The reason for this discrepancy is because we expect the
numerical value of $\mu$ to be larger than the lower bound given by
Eq.~(\ref{mu}).  Even though Eq.~(\ref{mu}) is not precise, it
captures the right qualitative behavior of $\mu$ with $\lambda$,
$\left< k \right>$ and $N$.  Indeed, we notice that the coupling
between the layers has no effect on $\mu$, as it is independent of $q$
and  $\omega$.  This is because we assume that the extinction in layer
A is not affected by layer B, as long as B is in the inactive state
(see Fig. \ref{fig:AB}).  We also observe that $\mu \to 0$ as $N \to
\infty$, in agreement with the fact that for $\lambda > \lambda_c$ an
infinite large system never decays to the healthy state.  That is,
once there is an outbreak in layer A, the transition back to the
healthy state is only possible in finite systems.

\section{Calculation of the outbreak rate $\beta$}
\label{betaapen}

To obtain the outbreak rate $\beta$ we run MC simulations on two
interconnected ER networks A and B, of $N=10^3$ nodes each, mean
degree $\left< k \right>=10$, $\lambda=0.115$ and coupling $q=0.005$,
starting from a density $I_A(0) = 0.11$ and $I_B(0)=0$ of infected
nodes in A and B, respectively.  We stop the simulation when either
$I_A$ becomes zero ($1 \to 0$ transition) or $I_B$ overcomes the value
$I^s=0.11$ by the first time ($1 \to 2$ transition).  This corresponds
to having the system initially in state $1$, with A active and B
inactive (see Fig.~\ref{fig:AB}), and calculating the first-passage
statistics to either state $0$ or $2$.  Given that $\beta$ is a
measure of the likelihood that the disease is transmitted from A to B,
we expect $\beta$ to decrease as the rewiring rate $\omega$ increases.
Therefore, we run simulations for several values of $\omega$ to study
this dependence.  To obtain the transition rates, it proves useful to
work with the persistence probability in state $1$, $S_1(t)$, i e.,
the probability that the system did not leave state $1$ up to time
$t$.  Given that the total outgoing rate from state $1$ is $\mu +
\beta$, we expect the persistence probability to behave as
\begin{equation}
S_1(t) \simeq e^{-(\mu+\beta)t}.
\label{S1}
\end{equation}

\begin{figure}[t]
\includegraphics[width=7.0cm]{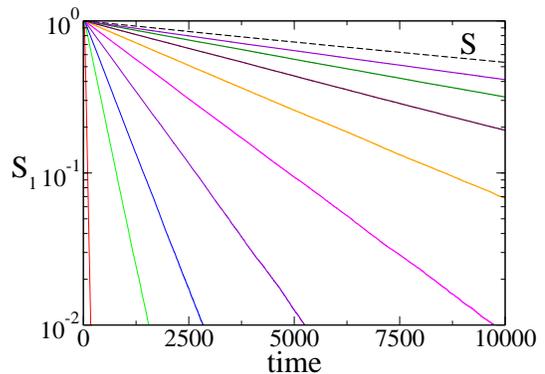}
\caption{Persistence probability in state $1$ $S_1$ vs time, for two
  coupled networks of $N=10^3$ nodes each, mean degree $\left< k
  \right>=10$, coupling $q=0.005$ and infection rate $\lambda=1$.
  Each solid curve corresponds to a different value of the rewiring
  rate $\omega=1280, 640, 320, 160, 80, 40, 20, 10$ and $0$ (from top
  to bottom).  The dashed curve at the top corresponds to the survival
  probability $S$ of a single network.  We note that the slope
  $\alpha$ of $S_1$ approaches the slope $\mu$ of $S$ as $\omega$
  increases.}
\label{fig:S1-t}
\end{figure}

As it happens for the single-network survival probability $S(t)$, we
found from MC simulations that $S_1(t)$ is nearly constant up to a
time $\hat{t} \simeq 50$, and then decays exponentially fast to zero
(see Fig.~\ref{fig:S1-t}):
\begin{eqnarray}
 S_1(t) \simeq \left\{ \begin{array}{ll} 1 & \mbox{for}~~t \le \hat{t}
   \\ e^{-\alpha(t-\hat{t})} & \mbox{for}~~t > \hat{t},
\end{array} \right.
\label{S1net}
\end{eqnarray}
where the exponent $\alpha$ depends on the rewiring rate $\omega$.
Again, it seems that after a time $\hat{t}$ the dynamics of the
two-network system behaves, at a coarse-grained level, as the one of a
three-state system with effective transition rates.  We observe in
Fig.~\ref{fig:S1-t} that $S_1$ approaches $S$ as $\omega$ increases.
This is consistent with the fact that at very large rewiring rates
network A decouples from network B.  Thus, A behaves as a single
network, independent from B and, therefore, the only possible process
is the extinction from the initial active state in A.

Comparing Eqs.~(\ref{S1}) and (\ref{S1net}) we obtain the relation
\begin{equation}
\alpha = \mu + \beta.
\end{equation}

\begin{figure}[t]
\includegraphics[width=7.0cm]{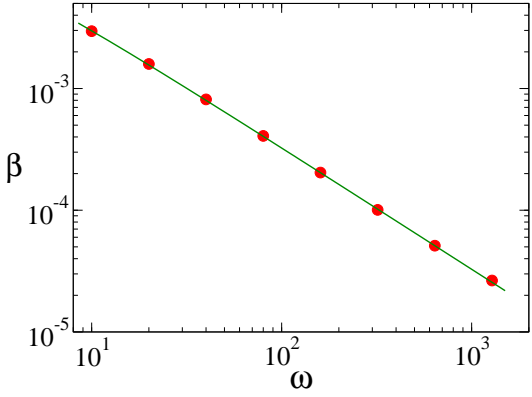}
\caption{Outbreak transition rate $\beta$ vs rewiring rate $\omega$.
  Solid circles were calculated from the slopes $\alpha$ of the
  persistent probability $S_1$ of Fig.~\ref{fig:S1-t}, using the
  formula $\beta=\alpha-\mu$, with the extinction rate $\mu=6.49
  \times 10^{-5}$.  The solid line corresponds to the fitting function
  $c/(1+\omega)$, where $c=0.0328$ is the best fitting parameter.}
\label{fig:beta-w}
\end{figure}

In Fig.~\ref{fig:beta-w} we plot $\beta=\alpha-\mu$ vs $\omega$, using
the measured values of $\alpha$ from Fig.~\ref{fig:S1-t} and the value
$\mu = 6.49 \times 10^{-5}$ calculated in Appendix \ref{muapen}.  The
data is well fitted by the function $\beta(w) = c/ (1+\omega)$, with
$c=0.0328$.  This means that $\beta$ vanishes as $\omega$ goes to
infinity, showing that the outbreak transition $1 \to 2$ is strictly
zero only in the $\omega \to \infty$ limit.

We now obtain an analytical expression for $\beta$.  Starting from an
initial condition consisting of an outbreak in network A and no
infected nodes in network B (state 1), we assume that a node in B gets
infected (``infection seed'') at a rate $\Lambda$, and that this seed
grows generating an outbreak with probability $S_{\infty}$.
Therefore, $\beta$ is estimated as
\begin{eqnarray}
\beta = \Lambda \, S_{\infty}.
\label{beta-1}
\end{eqnarray}
That is, every mean time interval $1/\Lambda$ a seed is planted in B,
which grows and becomes an outbreak with probability $S_{\infty}$, and
thus the outbreak probability per time is as in Eq.~(\ref{beta-1}).
Here
\begin{eqnarray*}
S_{\infty} \equiv \lim_{t \to \infty} S(t),
\end{eqnarray*}
where $S(t)$ is the survival probability in a spreading experiment, i
e., the probabilility that a run starting from a seed survives up to
time $t$.  Based on a rigorous proof for spreading models with
absorbing states \cite{Durrett:1988}, we expect the ultimate survival
probability to be equivalent to the density of nodes in the stationary
state,
\begin{eqnarray}
 S_{\infty} = I^s.
\label{Sinfty}
\end{eqnarray}
Besides, the total infection rate from A to B, $\Lambda$,  should be
proportional to the total number of interlinks of type $I_AS_B$ and
the infection rate per link $\lambda$, that is
\begin{eqnarray}
\Lambda = \lambda N I_AS_B.
\label{Lambda}
\end{eqnarray}
Here $I_AS_B$ is the stationary density of links connecting an
infected node in A and a susceptible node in B, which is estimated as
$I_AS_B=I_A^s \left< k_{AB} \right> = I_A^s \, q \left< k
\right>/2(1+\omega)$, given that all nodes in B are susceptible.  We
have also used the effective coupling $q_{\omega}=q/(1+\omega)$ for a
dynamic network, which accounts for the reduction of the number of
interlayer links between infected and susceptible nodes when the
rewiring is switched on.  Plugging this expression for $I_AS_B$ into
Eq.~(\ref{Lambda}) we obtain
\begin{eqnarray}
\Lambda = \frac{\lambda \, q \left< k \right> N I_A^s}{2(1+\omega)}.
\label{Lambda-1}
\end{eqnarray}
Finally, replacing the expressions for $S_{\infty}$ and $\Lambda$ from
Eqs.~(\ref{Sinfty}) and (\ref{Lambda-1}) into Eq.~(\ref{beta-1}) leads
to
\begin{equation}
\beta = \frac{\lambda \, q \left< k \right> N (I_A^s)^2}{2(1+\omega)}.
\label{beta-2}
\end{equation}
Using the values of the model's simulations $N=10^3$, $\left< k
\right>=10$, $\lambda=0.115$ and $I_A^s=0.11$ in Eq.~(\ref{beta-2}) we
get $\beta = c/(1+\omega)$, with $c \simeq 0.0348$, which is
comparable to the best fitting parameter $0.0328$ of
Fig.~\ref{fig:beta-w} (about $6 \%$ difference).  This shows that
Eq.~(\ref{beta-2}) is a good approximation for $\beta$, for the
parameters used in the simulation.

To complete the analytic expression for $\beta$ in terms of the
model's parameters, we insert into Eq.~(\ref{beta-2}) the analytical
expression Eq.~(\ref{Is}) for $I_A^s$ obtained in section \ref{prev},
and replace back $\lambda_c$ by $\left< k \right>^{-1}$.  We finally
arrive to
\begin{equation}
\beta \simeq \frac{\mathcal D \, q \, N }{1+\omega},
\label{beta}
\end{equation}
with
\begin{eqnarray}
\label{D-coef}
\mathcal D \equiv \frac{\lambda \, \left< k \right>}{2} \left(
\frac{1+\left< k \right>^2}{2}  + \frac{1}{\lambda} - \frac{1+\left< k
  \right>}{2} \sqrt{ \left( 1-\left< k \right> \right)^2 +
  \frac{4}{\lambda} }  \right). \nonumber \\
\end{eqnarray}

From Eq.~(\ref{beta}) we see that $\beta$ increases linearly with $q$
and $N$.  This happens because $\beta$ is proportional to the total
number of infections from nodes in A to nodes in B through interlayer
links, which are proportional to $q$ and $N$.  In particular, A has no
effect on B ($\beta=0$) when $q=0$.  We also observe that $\beta$
decreases monotonically with $\omega$, showing that the rewiring
reduces the transmission rate of the disease from A to B.

\bibliographystyle{apsrev}

\bibliography{ref}

\end{document}